\documentclass{aa}
\usepackage{hyperref}
\usepackage{graphicx}
\usepackage{longtable,lscape}  
\usepackage{rotating}

\begin{document}

\title{Kinematics of the compact symmetric object OQ~208 revisited}

\author{
        Fang~Wu\inst{1,2,3}
   \and Tao~An\inst{1,3}
   \and Willem~A.~Baan \inst{4}
   \and Xiao-Yu~Hong\inst{1,3}
   \and Carlo~Stanghellini \inst{5}
   \and Sandor~Frey \inst{6}
   \and Hai-Guang Xu \inst{7}
   \and Xiang Liu \inst{8,3}
   \and Jingying Wang \inst{7}
   }
\institute{Shanghai Astronomical Observatory, Chinese Academy of Sciences, 200030 Shanghai, China
\and
Graduate University of the Chinese Academy of Sciences, 100049 Beijing, China
\and
Key Laboratory of Radio Astronomy, Chinese Academy of Sciences, 210008 Nanjing, China
\and-
Netherlands Institute for Radio Astronomy (ASTRON), Postbus~2, 7990\,AA Dwingeloo, The Netherlands
\and
Istituto di Radioastronomia - INAF, via Gobetti 101, 40129 Bologna, Italy
\and 
F\"{O}MI Satellite Geodetic Observatory, PO Box 585, H-1592 Budapest Hungary
\and 
Department of Physics, Shanghai Jiao Tong University, Shanghai, 200240, China
\and 
Xinjiang Astronomical Observatory, Chinese Academy of Sciences, Urumqi 830011, China
}
   \offprints{T.~An; antao@shao.ac.cn}
  \date{Received ..., 2012; accepted ..., 2012}

\abstract
{}
{
A long-timeline kinematic study of the archetypal compact symmetric object (CSO) OQ~208 sheds light on the physical properties of the most compact radio sources.
}
{
Archival data from the Very Long Baseline Array (VLBA) at 15 GHz over a time span of 13.6 yr were used to investigate the kinematics of the radio source. 
The flux density monitoring data obtained at the Michigan 26-meter radio telescope were also used as supplementary information for analyzing the geometry of the radio structure. 
}
{
At 2.3-GHz, the radio emission is dominated by two mini-lobes separated by $\sim$10 pc in a northeast-southwest (NE--SW) direction. At 8.4 and 15 GHz, each lobe  is further resolved into two subcomponents, which are identified as hotspots.
A knotty jet is linked with the NE hotspot and traces back toward the geometric center. The core is too weak to be detected.
Significant flux density variation is found in the primary hotspots with a maximum level of 62\% (NE1) and 19\% (SW1).
The flare epoch of NE1 is earlier than that of SW1 by approximately 5.00 yr, suggesting that the northeast lobe is advancing and the southwest lobe is receding. 
This light travel difference indicates a radial distance difference between the two hotspots of 1.53 pc, which indicates an inclination angle of about 80.8 degrees between the radio jet and the line of sight.
The angular separation rate between NE1 and SW1 is 0.027  mas yr$^{-1}$, corresponding to a projected speed of 0.133 c.
The inner jet knot (J1) moves at 0.047 mas yr$^{-1}$ (or 0.230 c), about 3.5 times the hotspot advancing speed.
}
{
The large viewing angle and the modest jet speed suggest a mildly relativistic jet. The jet axis is close to the plane of the sky. 
The separation rate and the distance between the two primary hotspots result in a kinematic age of $255 \pm 17$ yr, confirming that OQ~208 is indeed a young radio source.
In addition to the hotspot advancing motions, sideways motions provide evidence that the lobes are obstructed by the external interstellar medium. 
}

\authorrunning{Wu et al.}
\titlerunning{Kinematics of OQ~208 revisited}

\maketitle


\section{Introduction}

{\bf Compact symmetric objects} (CSOs) are a {\bf subclass} of extragalactic radio sources that are characterized by a compact double or triple radio structure {\bf with an} overall size less than 1 kpc. The physical nature of the compactness of CSOs is still a question under debate: the {\it youth} model 
(e.g. \cite{phi82,fan95,rea96}) proposes that 
CSOs are small because they are in the infant stage of the extragalactic radio source evolution; the {\it frustration} model (e.g. \cite{van84,odea91}) attributes the small size of CSOs to extremely strong confinement by the dense external medium. The two models define {\bf distinctly} different evolutionary {\bf fates} of CSOs. In the {\it youth} model, all young CSOs eventually evolve into large-scale double sources, {\it i.e.}, Fanaroff-Riley (FR, \cite{FR74}) type-II sources, over a few million years; whereas according to the {\it frustration} model, CSOs are confined within the host galaxy and experience stagnated growth. The CSO ages provide a critical {\bf distinction} between the two models. Previous kinematics studies of individual CSOs and subsamples of {\bf these} sources (e.g. \cite{oc98,ocp98,tay00,gug05,an12a}) show that the measured hotspot advancing speed values have a large scatter from $\sim0.04\,c$ to $\sim0.5\,c$, resulting in young kinematic ages in the range of only 100--2000 yr. However, additional sideways motions of hotspots and disturbed lobes indicate strong interactions between the jet heads and the surrounding interstellar medium (\cite{sta09,an12a}). Sideways motion is usually very slow compared to the dominant advancing motion, {\bf which requires} highly accurate measurements. A detailed investigation of well-selected CSO samples is essential for exploring the complex kinematics of CSOs, and for understanding the physical environment of the host galaxies on 1--500 pc scales.

The kinematic age ($\tau_{k}$) of CSOs is traditionally determined by dividing the separation ($R$) between two terminal hotspots by the separation rate ($\mu$). A {\bf high-accuracy} measurement of $\tau_{k}$ requires high-precision position determination for individual observations, and a sufficiently long timespan of the data. For more than two epochs, $\mu$ is often determined from a linear regression fit to the changing hotspot separation with time. The statistical uncertainty of the fit is sensitive to the number of available data points and the uniformity of the time sampling. 
In addition to {\bf this}, different resolutions of the interferometric images and {\bf the} intrinsic opacity effect in different levels may introduce systematic errors in the proper motion measurements. 
An accurate measurement for the separation rate $\mu$ requires CSOs with multiple-epoch Very Long Baseline Interferometry (VLBI) imaging data at the same observing frequency over a sufficiently long time span.

In this paper, we carry out a kinematic study of an archetypal CSO, OQ~208, on the basis of archival VLBI data over a time baseline of 13.6 yr.
\object{OQ~208} (also known as \object{Mrk~668}, \object{B1404+286}) is one of the closest  CSOs ($z = 0.0766$: \cite{huc90}) and provides a template for the dynamic properties of the most compact (and possibly the youngest) radio sources. The host galaxy of OQ~208 shows typical Seyfert-1 spectra with strong broad Balmer lines (FWHM$_{H\alpha}=6000$ km s$^{-1}$) and also forbidden lines of  [Ne III], [O III] and [S II]  (\cite{BS72,EH94}). In the radio band, a broad 21-cm H{\sc I} absorption line was detected in OQ~208, indicating a fast and massive outflow of neutral gas (\cite{Mor05}). The radio emission is concentrated in a compact central region within $\sim$10 mas (\cite{sta97}). Additional evidence for the compactness comes from the convex radio spectrum with a turnover at about 4.9 GHz (\cite{dal00}), resulting from synchrotron self-absorption or free-free absorption (\cite{fan09}). The VLBI images of OQ~208 at 2.3 and 5 GHz reveal a typical CSO-type morphology with two compact (mini-)lobes along the NE--SW direction (\cite{sta97,liu00,wang03}). At higher frequencies of 8 and 15 GHz, a more detailed structure is revealed: the mini-lobes are resolved into subcomponents, and internal jet knots are detected between the two lobes (\cite{kel98,luo07}) Previous proper motion measurements of OQ~208 (e.g. \cite{liu00,sta02,wang03,luo07}) were mostly based on 5- and 8-GHz data with lower angular resolutions and {\bf are} affected by the opacity effect. Moreover, these measurements only cover a time span of a few years, insufficient for tracing the long-term change of the hotspots. In the present study, we only {\bf made} use of 15-GHz VLBI data to determine the separation velocities of hotspots and the proper motions of the internal jet knots.

The structure of the paper is as follows. Section \ref{sec2} describes the VLBI data used for the kinematics study. Section \ref{sec3} presents the results, including radio images, light curves, and proper motion measurements. A conclusion and a summary are given in Section \ref{sec4}. Throughout this paper, we assume a flat cosmological model with $H_{0}$ = 73 km s$^{-1}$ Mpc$^{-1}$, $\Omega_{M} = 0.27$, $\Omega_{\Lambda} = 0.73$. At the redshift of OQ~208, 1 mas angular size corresponds to 1.4 pc projected linear size, and a proper motion of 1 mas yr$^{-1}$ corresponds to 4.9 $c$ apparent speed.

\section{VLBI data} \label{sec2}

{\bf Owing} to its compactness and high brightness, OQ~208 is often used as a calibrator in VLBI experiments. Consequently, there are abundant archival VLBI data sets from the past two decades.
The 15-GHz VLBA data from the Monitoring of Jets in Active Galaxies with VLBA Experiments (MOJAVE) program\footnote{http://www.physics.purdue.edu/MOJAVE/} have the highest resolution and sensitivity and are most appropriate for kinematic studies. The 15-GHz data spread over 28 epochs in the time range from early 1995 to the middle of 2009.
We also made use of the VLBI data from the Radio Reference Frame Image Database (RRFID)\footnote{http://rorf.usno.navy.mil/rrfid.shtml}, which were obtained with the VLBA together with several geodetic radio telescopes simultaneously at 2.3 and 8.4 GHz. The multiple-frequency data were used for the spectral index analysis of compact components.
The RRFID data cover the time range from 1994 to 2008.

\begin{figure*}
\centering
\includegraphics[angle=0,width=0.45\textwidth]{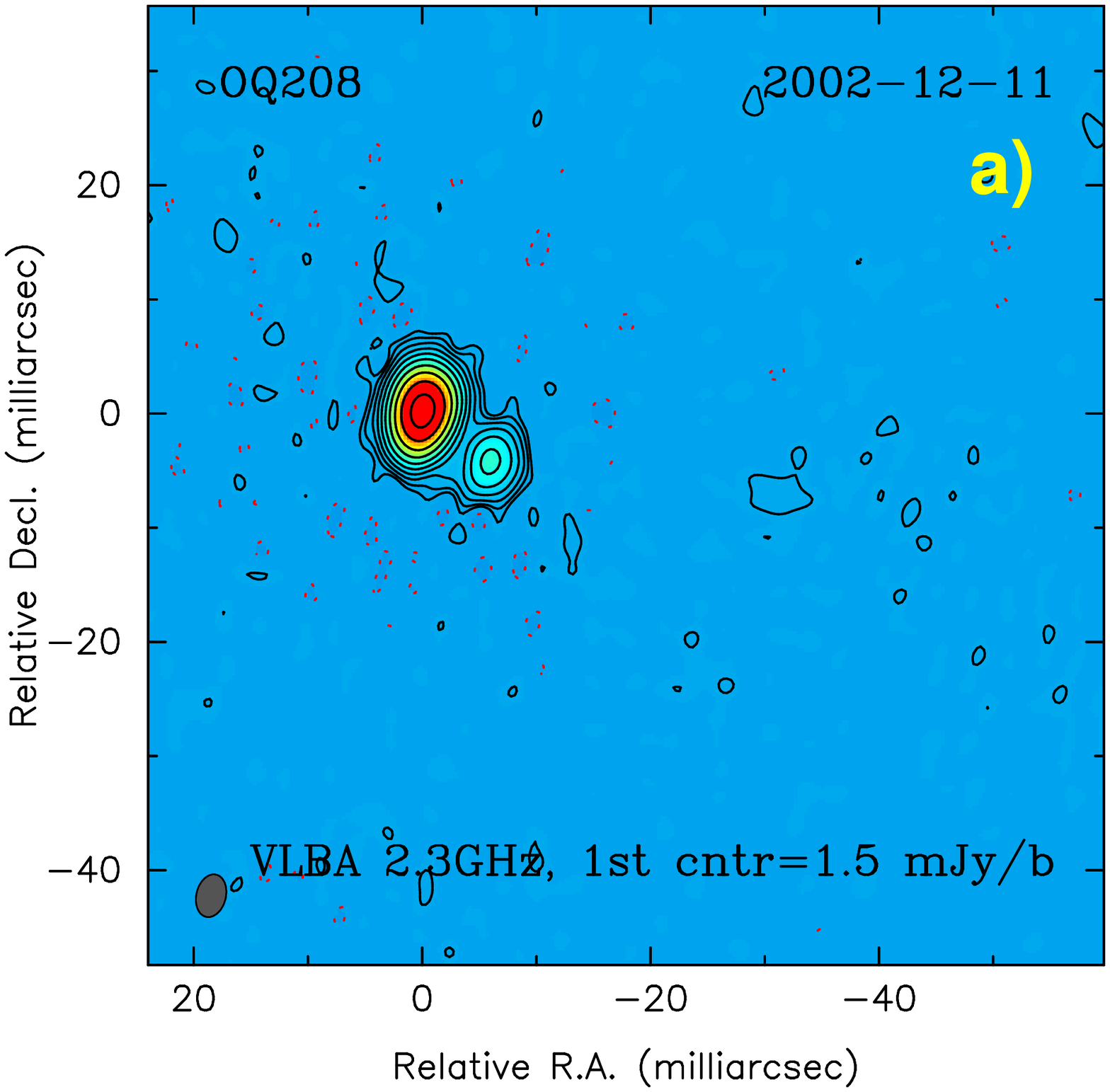}
\includegraphics[angle=0,width=0.45\textwidth]{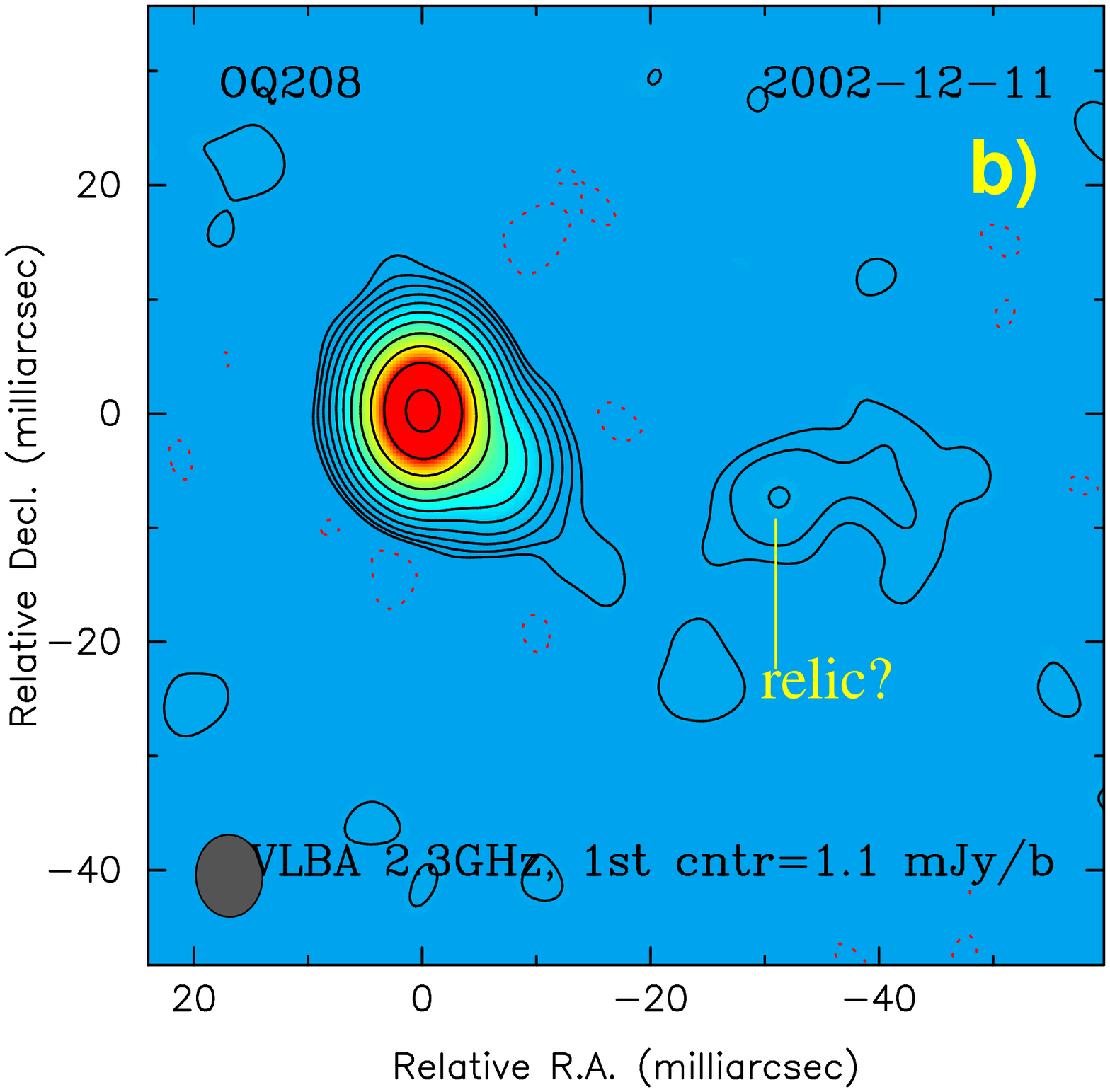}
\includegraphics[angle=0,width=0.45\textwidth]{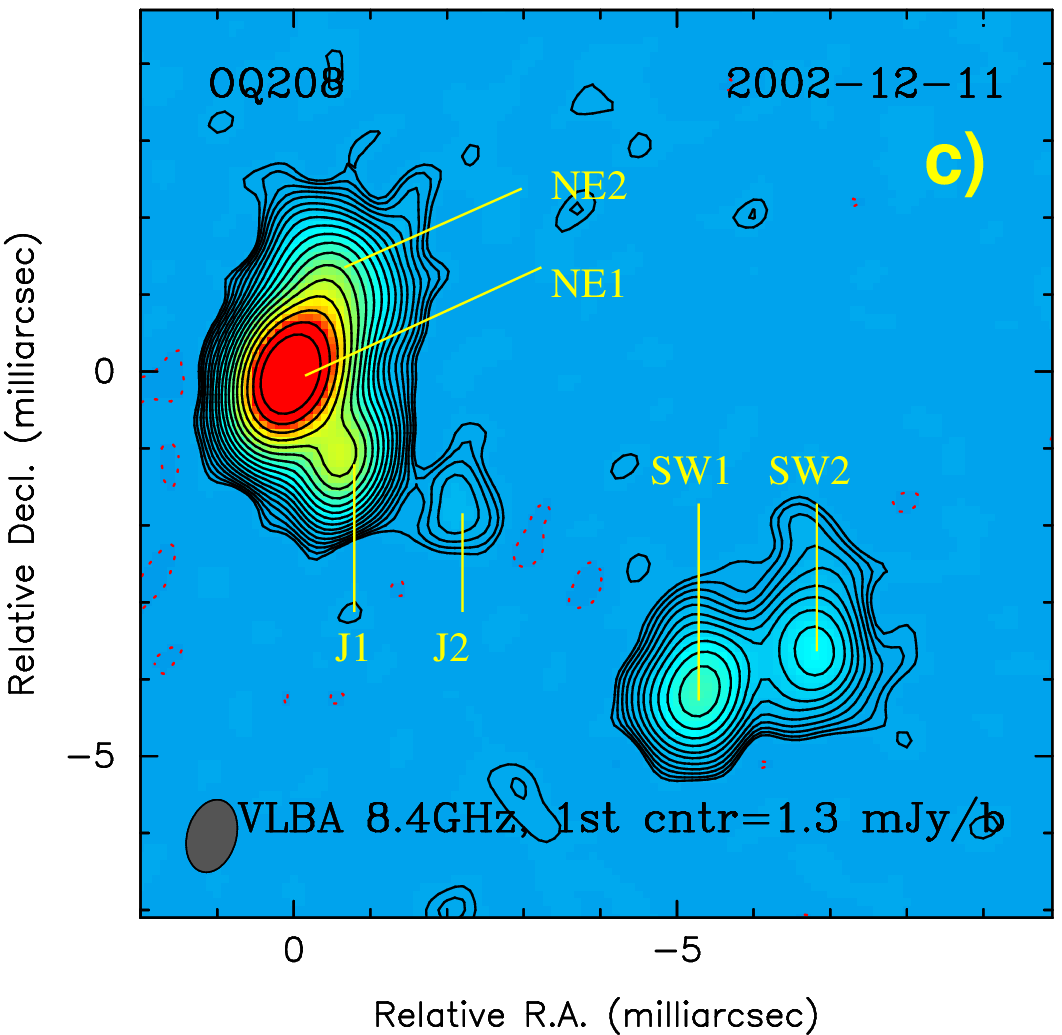}
\includegraphics[angle=0,width=0.45\textwidth]{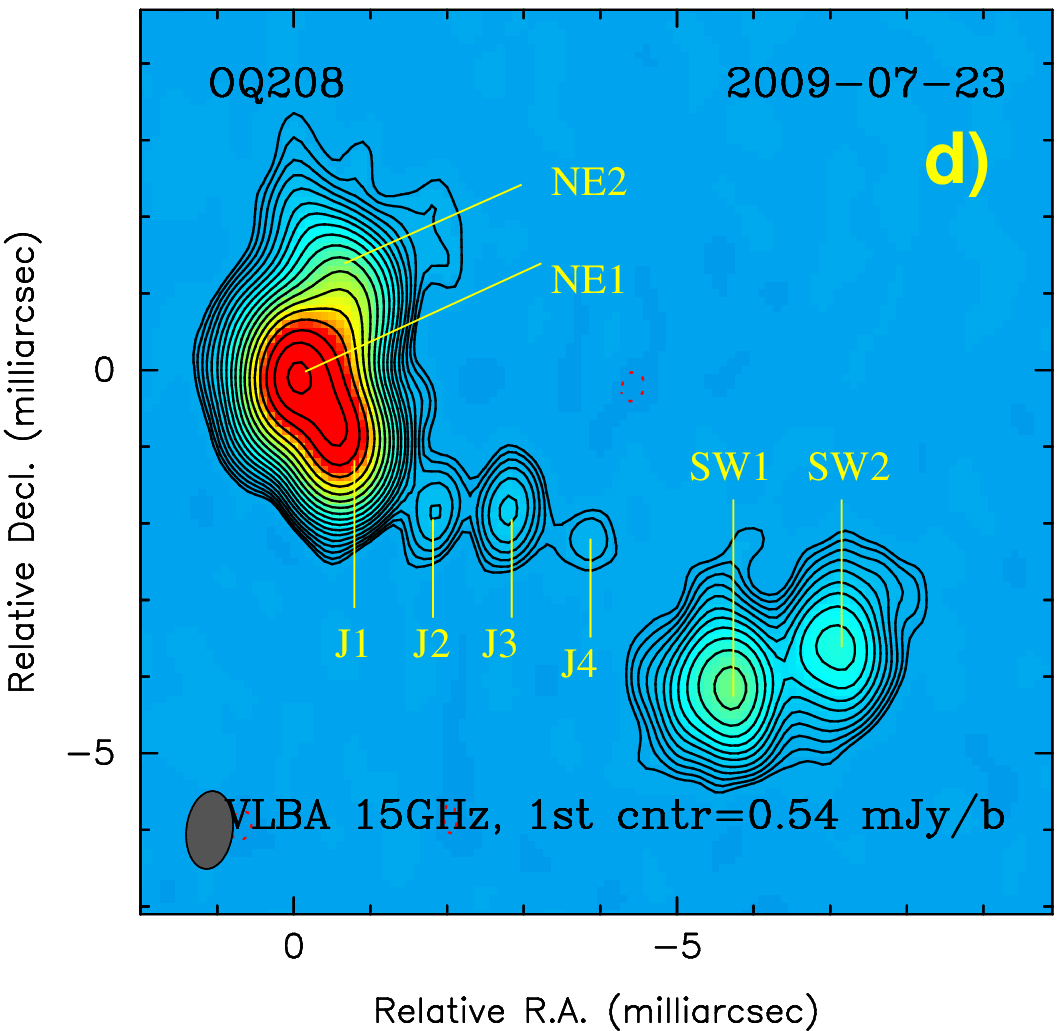}
\caption{{\bf Total} intensity image of OQ~208. Panel {\it a} shows the 2.3-GHz VLBI image, displaying a double-component morphology. Panel {\it b} shows the tapered 2.3-GHz image. The extended emission at $\sim$30 mas west of the primary structure is likely a relic radio emission resulting from the past activity a few thousand years ago. Panels {\it c} and {\it d} show the 8.4 and 15 GHz images. Each mini-lobe is resolved into two hotspots. Between the hotspots NE1 and SW1, a knotty jet is detected.}
\label{fig:mor}
\end{figure*}

The calibration of the archival VLBI data has already been made in the Astronomical Image Processing System (AIPS) following the standard procedure. We performed {\bf additionally several} iterations of self-calibration in DIFMAP (\cite{she94}) to eliminate residual antenna-based phase errors. The 15-GHz visibilities were fitted with five Gaussian model components (two in the northwest lobe, another two in the southeast lobe, and one jet knot, see Figure \ref{fig:mor}) using the DIFMAP task MODELFIT. Table 1 lists the fitted parameters, including the integrated flux density S$_{i}$, the relative separation $R$ from the primary hotspot used as the reference, the position angle of the VLBI component, and the deconvolved size $\theta_{maj}\times\theta_{min}$. The statistical uncertainties of the observed parameters were estimated according to the formulae given by \cite{fom99}, except that we also included an additional 5\% as the amplitude calibration error. The details of the error analysis method are given by \cite{an12a}.

In addition to VLBI imaging data, the single-dish flux density monitoring data observed with the 26-meter paraboloid telescope of the University of Michigan Radio Astronomical Observatory\footnote{https://dept.astro.lsa.umich.edu/datasets/umrao.php} (\cite{all85}) were included for a supplementary analysis of the variability and the geometry of the radio structure. The total flux density measurements were made at 4.8, 8, and 14.5~GHz from July 1974 to August 2009.

\section{Results} \label{sec3}

\subsection{Radio morphology}
\label{sec:mor}
At 2.3 GHz, OQ~208 exhibits the double-component morphology typical of the CSOs (Figure~\ref{fig:mor}-a). The total extent of the radio source is about 7 mas ($\sim$10 pc). The northeast (NE) component is much brighter than the southwest (SW) one with an intensity ratio of 18.0:1 at 2.3 GHz. 
Both components show a steep spectral index at high frequencies
(defined as $S\propto \nu^{-\alpha}$) with $\alpha_{8.4GHz}^{15GHz}= 1.38 \pm 0.03$ (SW) and $\alpha_{8.4GHz}^{15GHz}= 1.12 \pm 0.02$ (NE), identifying them as two mini-lobes of the CSO.
The asymmetric brightness of two-sided jets/lobes in radio sources is commonly attributed to the Doppler boosting effect that enhances the apparent flux density of the advancing jet/lobe by a factor of $\delta^{3+\alpha}$, where $\delta$ is the Doppler boosting factor and $\alpha$ is the spectral index. However, the kinematic analysis below suggests that the jet in OQ~208 is mildly relativistic 
and the jets nearly align with the plane of the sky, therefore Doppler boosting cannot account for the brightness difference. 
Another explanation of the {\bf strong} asymmetry in the brightness of the two lobes is that the receding SW lobe suffers from more free-free absorption than the advancing NE lobe. Indeed, the flux density ratio is even {\bf higher} at a lower frequency of 1.66 GHz, $S_{NE}/S_{SW}=60:1$ (\cite{kam00}), supporting this interpretation. 
A third possibility is that the inhomogeneous distribution of the external medium surrounding the lobes results in a larger  conversion efficiency from jet kinetic energy to radiative energy in the northeast jet (\cite{ori12}). 

A 2.3-GHz image with a Gaussian taper with a half-value at 20~M$\lambda$ wavelength (Figure~\ref{fig:mor}-b) reveals  an extended feature about 30 mas ($\sim$40 pc) to the west, which was previously reported by \cite{luo07}. Extended emission features on kpc to Mpc scales has also been detected in OQ~208 (\cite{deB90}) and in other CSO galaxies (e.g., 0108+388: \cite{bau90,sta05}; 0941-080, 1345+125: \cite{sta05}), which were interpreted as relics remaining from past ($>10^8$ yr ago) nuclear activity. Extended components on scales of $<$100 pc are rarely seen in CSOs (J1511+0518 is {\bf another} example; \cite{ori08b}) and this puzzling one-sided extended feature at 40 pc distance requires an interval between two intermittent activities shorter than $2\times10^3$ yr (\cite{ori08b}).  The non-detection of {\bf the} northeast fading lobe likely indicates asymmetric properties of the ambient {\bf interstellar medium} (ISM) on pc scales, leading to more rapid radiative or adiabatic losses and a shorter life of the NE lobe. This, again, is consistent with the fact that the NE advancing lobe is much brighter.

The NE lobe is resolved into two subcomponents (NE1 and NE2) at 8 and 15 GHz. NE1 dominates the flux density of the whole source. The spectral index of NE1, determined from 8 and 15 GHz data at close epochs, is $\alpha^{15GHz}_{8GHz}=0.99 \pm 0.18$. The brightness temperature is calculated using the equation
   \begin{equation}\label{eq:Tb}
    T_{b} = 1.22 \times 10^{12}\frac{S_{ob}}{\nu_{ob}^{2}\theta_{maj}\theta_{min}}(1+z),
  \end{equation}
  where $S_{ob}$ is the observed flux density in Jy, $\nu_{ob}$ is the observing frequency in GHz, $\theta_{maj}$ and $\theta_{min}$ are the major and minor axis of the Gaussian model component in units of mas, and $T_b$ is the derived brightness temperature in source rest frame in units of Kelvin. The average brightness temperature of NE1 is $4.4 \times 10^{10}$ K.
The secondary component NE2 is weaker than NE1 in the range 4.3--16.3. It has a lower brightness temperature of $1.4 \times 10^{9}$ K and a much steeper spectral index of $\alpha^{15GHz}_{8GHz}=2.09 \pm 0.13$.
The high brightness temperature and the relatively flatter spectral index of NE1 identify it as the primary hotspot formed by the reverse shock when the jet head impacts on the wall of the external medium.
The continuous emission structure between NE1 and NE2 would suggest a physical connection.

The SW lobe is resolved into two components (SW1 and SW2) at 8 and 15 GHz. The two components have comparable sizes. The integrated flux density of SW1 is slightly higher than SW2 with a flux density ratio $R(SW1/SW2)$ in the range of 1.3--2.4. 
Their spectral indices are $\alpha^{15GH}_{8GHz}=1.11\pm0.07$ (SW1) and $1.82\pm0.26$ (SW2).

The presence of double hotspots in the 10-pc lobe structure of OQ~208 remains a puzzle. Similar morphology is found in {\bf the} multiple-hotspot appearance of large-scale double sources such as 3C~20 (\cite{lai81,har01}), where the hotspot usually identified as the primary hotspot is extremely compact and bright, and the other is more diffuse and shows various structures. 
Three models may account for the double hotspots inside the 10-pc nuclear region of OQ208:

(1) {\it Blocking of the jet head by the external medium}. In this scenario, the jet impacts on the wall of the surrounding medium and generates the hotspot there. As the jet head may slide along the wall, the hotspot position also changes accordingly (dentist's drill model: \cite{sch82}). The primary hotspot is the {\bf currently} active jet-ISM interaction, while the secondary hotspot represents the past primary hotspot, which is now fading away. 

(2) {\it Jet precession}. The present images do not allow us to distinguish whether a single jet beam is precessing (\cite{cox91}) or {\bf if} there are intrinsically two beams (\cite{lai82}) due to the similar compactness and shape of two hotspots and the absence of emission between the hotspots and the nucleus. 
One possible driving mechanism of jet precession is associated with a binary black hole system where the black hole launching the jet has a relative motion with respect to the surrounding ISM and the multiple hotspots reflect the different impact locations. 
Taking the brighter SW1 as the primary (current) and SW2 as the secondary (older) impact of the jet beam(s) on the external medium, the older hotspot has stopped receiving direct energy supply from the nucleus, and synchrotron aging will result in a steeper radio spectrum at high frequencies. Indeed, SW2 has a steeper spectral index than SW1. If SW2 is the old hotspot, it should lie behind the primary hotspot SW1, implying {\bf that} there is a jet bending between SW2 and SW1.  

(3) {\it Redirected flows}. Alternatively, the secondary hotspot represents either a re-directed outflow originating from the primary hotspot (\cite{lai81}) or a deflected jet flow (\cite{LB86}) escaping from a weak point in the cocoon/lobe structure.
Different from the model (1), the secondary hotspot in model (3) continues to receive energy supply from the primary hotspot. Therefore, it would last longer, would not show spectral steepening, and be much fainter and less compact than the primary hotspot.
In the eastern lobe, NE2 has no direct connection with the nucleus, but shows a collimated bridge {\bf that links it} with NE1. NE2 is also larger than NE1 and has an amorphous shape.
These morphological characters make NE2 more likely {\bf to be} the result of a re-directed outflow from the primary hotspot NE1 than a directed flow from the nucleus.

The continuous emission between two hotspots NE1 and SW1 at 2.3-GHz (Figure~\ref{fig:mor}-a) is resolved into a knotty jet at 8.4 and 15 GHz (Figure~\ref{fig:mor}-c and \ref{fig:mor}-d). From east to west, these knots are labeled J1, J2, J3, and J4. J1 is the brightest jet knot located at the inlet of the northeast lobe and is detected at 15 and 8.4 GHz at all epochs. It has a 15 GHz to 8.4 GHz spectral index between 0.81 and 1.45, indicative of a typical optically thin jet knot, and a brightness temperature of $3.4 \times 10^9$~K. J2 appears in the 8.4-GHz images at epochs from 2002.044 to 2002.945, where the VLBI images have the highest sensitivity and resolution. J2 only appears in the most sensitive 15-GHz image at epochs 1995.958, 1996.378, 2003.318, 2008137, and 2009.558. It has a steep spectrum with $\alpha^{15GHz}_{8GHz}=1.43$ and a brightness temperature $T_{b} \leq  1.9 \times 10^7$~K. J3 is closest to the geometric center and only appears in the highest sensitivity 15-GHz images.  The brightness temperature of J3 is $ \leq 3.0\times 10^{7}$~K derived from the 15-GHz data. Lacking the spectral index information, the active nucleus or the innermost jet knot nature of J3 remains uncertain. Regardless of the identification of J3, the non-detection at 8.4 GHz would imply an extremely high absorption opacity toward the nucleus. J4 is only marginally detected at 15 GHz on 2009.558. More sensitive observations are necessary to to reveal the physical nature of these components.

\subsection{Variability of the hotspot NE1}

\begin{figure*}
\centering
\includegraphics[angle=0,width=0.95\textwidth]{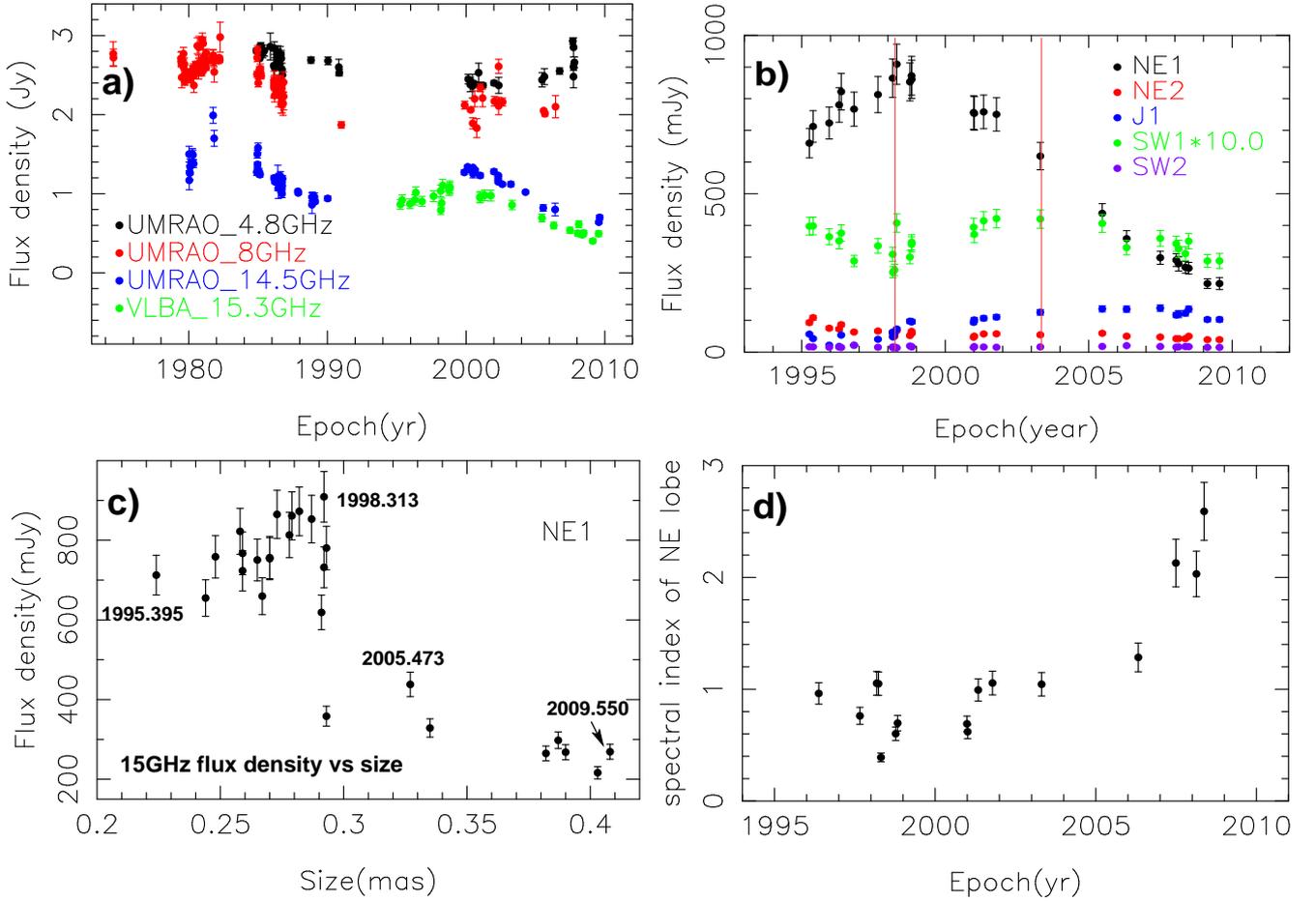} 
\caption{Variability of VLBI components of OQ~208.
(a): Total flux density variability of OQ~208 observed with the UMRAO 26-meter radio telescope: 4.8 GHz (black), 8 GHz (red), 14.5 GHz (blue). The sum of 15-GHz flux densities of the VLBI components (see Table 1) is represented by green symbols indicating that $\sim$15\% is missed due to lack of short uv-spacings;
(b): Temporal variations of the flux densities of VLBI components. The flux densities are derived from the model-fitting results listed in Table 1. Two vertical lines mark the peak epochs of NE1 (around 1998.313) and SW1 (around 2003.313). To clearly show the variability behavior of SW1, the flux densities of SW1 have been multiplied by a factor of 10;
(c): Variation of the 15-GHz flux density of the hotspot NE1 with the component size. The flux density of NE1 experienced a slow rise before epoch 1998.313, followed again by a sharp decline until epoch 2005.473 followed by a slow decline. The component size constantly increases during the whole period;
(d): Temporal variation of the spectral index $\alpha^{15GHz}_{8GHz}$ of the NE lobe. The spectral indices are calculated from the 15-GHz MOJAVE data and 8-GHz RRFID data at close epochs. A sharp jump of the spectral index occurs around 2006, after which the spectral index steepens significantly.}

\label{fig:lc}
\end{figure*}

The steep-spectrum lobes dominate the total flux density of most CSOs. \cite{FT01} monitored a sample of seven CSOs using the VLA at 8.5 GHz and found extremely stable flux densities (rms variability {\bf less} than 1\%) over a period of eight months.
Some exceptional CSOs show moderate-level variability on time scales of a few years, for example, OQ~208 (the present paper) and J1324+4048 (\cite{an12a}).
Unlike with blazars with a jet luminosity amplified by Doppler boosting, the low- and moderate-level variability observed in lobe-dominated CSOs is probably related to variation of the energy supply by the relativistic jet flow and the changes of the energy dissipation process including the adiabatic expansion losses and radiative losses.

The light curves observed with the Michigan 26-meter radio telescope (UMRAO) at three frequencies (4.8, 8, and 14.5 GHz) show two distinct flares in the past 25 years, around 1983 and 2000 (Figure~\ref{fig:lc}-a). The accurate peak epoch around 1983 cannot be clearly determined because of the broad gap in the sampling, while the second-peak epoch is not well constrained because the light curves only cover the declining part of the flare. 
The 5-GHz flux density variation of OQ~208 using the VLA \cite{sta97} shows a continuous decrease from 1980 to 1995.
In addition, Waltman et al. (1991) observed a decrease of the 8.1-GHz flux density from about 2.7 Jy in mid-1983 to about 2.0 Jy in mid-1989 using the Green Bank Interferometer. Starting from 1989, the flux density has been constant around 1.9 Jy. 
This evidence consistently suggests that OQ~208 has intrinsic variability on time scales of a few years.

For completeness, the 15-GHz summed flux densities of the VLBI components {\bf were} added to diagrams covering the time range between July 1995 and August 2009, which confirms the second flare peak between 1998 and 2000. The declining part of the VLBI flux density data is fully consistent with the total flux density data. The comparison of the VLBI and single-dish data also suggests that a significant {\bf percentage} ($>85\%$) of the total flux density originates from the compact radio structure. 

{\bf To} quantitatively evaluate the variability, we introduced a variability index $V$ to illustrate the relative variability scale (\cite{zha12}):
   \begin{equation}\label{eq:var}
    V = \frac{S_{max} - S_{min}}{S_{max} + S_{min}},
   \end{equation}
   where $S_{max}$, $S_{min}$ are the maximum and minimum flux density of an individual flare. A value of $V \approx 0$ represents indistinguishable or undetectable variability, and $V \approx 1$ indicates an extreme variability 
The 15-GHz light curves of individual components (Figure~\ref{fig:lc}-b) suggest a high-variability index $V = 0.62\pm0.12 $ for NE1 during the 1998 flare and a mild $V=0.19\pm0.04 $ for SW1 during the 2003 flare. The standard deviation of the measured flux densities with respect to the mean value is adopted as the statistical error. The variability index of NE1 derived from VLBI data is consistent with that measured from the single-dish monitoring data (Figure~\ref{fig:lc}-a), confirming that the total flux density variation is dominated by the brightest hotspot NE1. The other VLBI components do not show significant variability at 15 GHz. 

The variability at 8 and 14.5 GHz using the single-dish data (Figure~\ref{fig:lc}-a) indicates a {\bf maximum} level of $(22\pm4)\%$ and $(40\pm8)\%$, respectively. The 5-GHz variability scale is lower than at 8 and 14.5 GHz, probably {\bf because of} the increasing opacity at lower frequencies. 
The mini-lobes at 2.3 GHz based on the RRFID data did not show {\bf any} prominent variability. 
At 1.4 GHz, the total flux density of OQ~208 increased from $\sim$755 mJy in 1979 to $\sim$854 mJy in 2001 (de Bruyn, private communication), which is closely connected with the growth of the overall source size. 

The complex variability behavior of OQ~208 can be understood separately in the low- and high-frequency regimes: 

(1) Up to 1 GHz, the emission mostly arises from the extended structure surrounding the compact jets and lobes, which are strongly absorbed. 
The flux density of an opaque source can be simply expressed as $S_\nu \propto T_{b} \times A$, where $S_\nu$ is the flux density at an optically thick frequency $\nu$, $T_{b}$ is the brightness temperature at frequency $\nu$, and $A$ is the surface area of the extended structure. The continuous increase of the flux density from 1.4 GHz to 335 MHz results from the (slow) growth of the overall source size as a result of the expansion of the radio jets and lobes. 

(2) Above 1.4 GHz, the synchrotron radiation from the extended structure drops abruptly and the flux density is dominated by the compact hotspot, the mini-lobe, and the jet components.
Arguments can be made that radiative losses play a dominant role in the energy balance of  young and compact radio sources in the presence of strong (milligauss; mG) magnetic fields (\cite{sta97,ori06,ori12}). The adiabatic expansion would then only lead to a decrease in the optically thin section of the spectrum. 
However, the source does not show prominent variability at 2.3-GHz, which could mean that the source is still opaque at this frequency and that any low-amplitude flux density variation is smeared out by the opacity effect. 

(3) At frequencies above 5 GHz, the spectrum becomes optically thin and the observed variability is governed at the hotspots, i.e., the flux density from the most active jet-ISM interaction interface varies with the balance between the feeding and acceleration of fresh relativistic electrons and the radiative losses. 
The variation of the 15-GHz flux density with the component size (Figure~\ref{fig:lc}-c) shows and increase of the flux density increases with the increasing size before the flare peak at epoch 1998.313, indicating the injection of particles, energy, and magnetic fields into the lobes. During this stage, the hotspot is still optically thick. 
After 1998, the intermittent feeding through the jet decreases and the hotspot continues to expand adiabatically and to become more optically thin. The strong synchrotron radiation results in a sharp decrease of the flux density from $\sim$900 mJy at 1998.313 to $\sim$400 mJy at 2005.473. Figure~\ref{fig:lc}-d shows the variation of the spectral index $\alpha^{15GHz}_{8GHz}$ with time, using the 15-GHz MOJAVE data and 8.4-GHz RRFID data at close epochs. After the clear division around epoch 2006, the spectrum is significantly steepened. This supports the {\bf idea described above} that the flux density variation of hotspot NE1 is regulated by intermittent nuclear feeding. When there are no or less fresh relativistic electrons inserted into the hotspots, the old electrons suffer from synchrotron aging, and the spectral index in the optically-thin part steepens. 

Stanghellini et al. (1997) estimated the lifetime of the electrons radiating at 5 GHz {\bf assuming} equipartition magnetic fields and a homogeneous spherical geometry for the hotspot. The resulting magnetic field in hotspot NE1 (component A in Stanghellini et al. 1997) is 90 mG, and an electron lifetime at 5 GHz of 25 yr. The synchrotron aging time scales with the observing frequency and the brightness temperature $T_b$ as $t_{syn} \sim \nu^{-2} T_b^{3}$. Using the same arguments, the lifetime of electrons radiating at 15 GHz is about 4 yr, in agreement with the observed variability time scale of $\sim$15 yr (from 1983 to 1998).

\subsection{Geometry of the radio structure}

According to Figure~\ref{fig:lc}-b, the light curve of the northeast hotspot NE1 peaks around 1998.313. The peak of the light curve of SW1 is not very prominent and falls in a broad time range between 2002 and 2005. A mediate epoch of $2003.31^{+2.16}_{-1.53}$ is chosen as an estimate of the peak epoch, in which the uncertainty is represented by the scattering of the peak epoch. If the radio flux density enhancement in both hotspots is associated with the same episodic nuclear activity, the observation that {\bf  the flare of NE1 is leading that of SW1} by $5.00^{+2.16}_{-1.53}$ yr suggests that NE1 is the advancing lobe and SW1 is the receding one. The light travel time along the line of sight between the two hotspots corresponds to a radial difference of $5.00^{+2.16}_{-1.53}$ light years, or $1.53^{+0.67}_{-0.47}$ pc. 
Assuming that the two lobes are symmetric relative to the core (separation $\sim$9.5 pc) and that both jets are ejected at the same velocity, an inclination angle of the radio structure of $80.8\degr^{+2.8\degr}_{-3.8\degr}$ may be derived between the jet and the line of sight.
This calculation suggests that the jet axis of OQ~208 is very close to the plane of the sky. It is supported by the observation that the jet emission at the nucleus is not Doppler boosted (Sections 3.1 and 3.4). 

However, this implication appears to {\bf be} conflict with the optical spectroscopic identification of OQ~208 as a Seyfert 1 galaxy with broad Balmer emission lines (\cite{BS72}). In the standard unification model for AGNs (\cite{ant93}), the Type-I AGNs should have an inclination angle less than 70\degr (\cite{zha05,liu07}). Such an angle of 70\degr would predict a time delay between NE1 and SW1 of about 10.97 yr, for which there is no evidence in the light curves. The inconsistency between these inclination angles may suggest that the radio jet does not align with the optical axis of symmetry. 

If the brightness asymmetry of the two lobes is {\bf caused by} differential free-free absorption (Section 3.1), the physical properties of the ISM can be constrained. The spectral fit of free-free absorption gives a differential opacity $\Delta \tau_{ff}=5.3$ (\cite{kam00}). The electron density is about 2700 cm$^{-3}$ along a 1.53 pc path, assuming that the ISM surrounding the northeast and southwest lobes is homogeneous and the electron temperature is $T_{e} = 10^4$ K.  This electron density is typical for the narrow-line regions (NLR). An inhomogeneous ISM would require a higher density (in a clumpy medium) to account for the free-free absorption discrepancy.

\subsection{Kinematics and jet properties -- a young radio source}

\begin{figure}
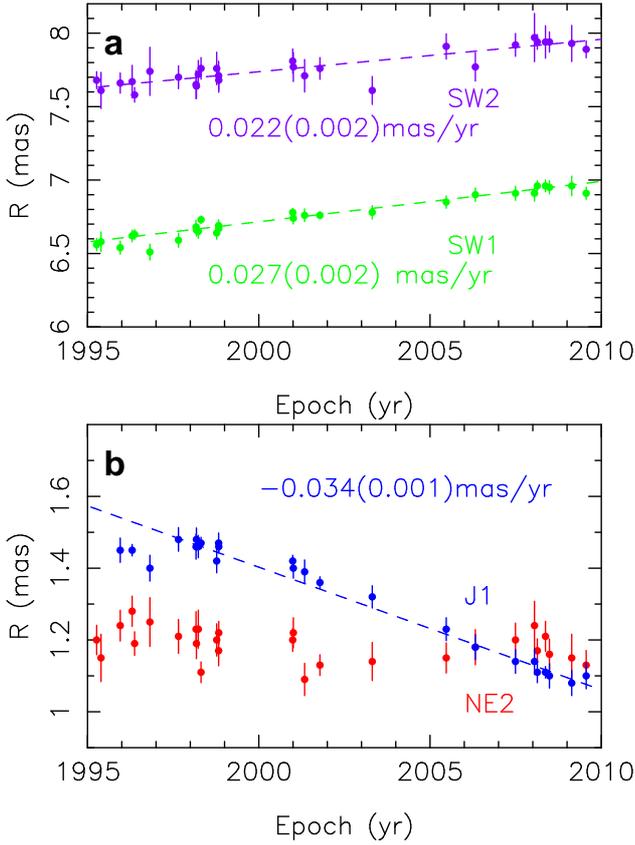

\centering
  \includegraphics[angle=-90,width=0.45\textwidth]{fig3a.ps} 
  \includegraphics[angle=-90,width=0.45\textwidth]{fig3b.ps}
  \caption{Relative proper motions of VLBI components with respect to the brightest hotspot NE1. The proper motions of the VLBI components are determined by linear fits to the separations as a function of time. The angular separation rate is calculated as $\mu(J1) = -0.034\pm0.001$ mas yr$^{-1}$,  $\mu(SW1) = 0.027\pm0.002$ mas yr$^{-1}$, and  $\mu(SW2) = 0.022\pm0.002$ mas yr$^{-1}$. The hotspot NE2 does not show significant proper motion.
}
  \label{fig:sep}
\end{figure}

Existing kinematics studies of CSOs show that the hotspot advancing speed is typically around 0.2$c$, and that they are young with kinematic ages of only between 100 and 2000 yr (\cite{oc98,ocp98,pc03,tay00,gug05,an12a}). 
In the present work, the slow proper motion of OQ~208 components are determined {\bf from} making use of the long time-baseline 15-GHz VLBA data to eliminate systematic errors induced by different resolutions, different {\it uv} coverage, and different frequencies. Since the identification of the central core of OQ~208 is not certain yet, the brightest hotspot NE1 is used as the reference, and relative proper motions of other VLBI components are measured.  

Figure~\ref{fig:sep} shows the variation of the distance of the VLBI components using 28 individual measurements over a time span is 13.6 yr.
Linear regression was used to determine the separation rate. The positional uncertainty ($\sigma \approx 0.02$ has) of each component at each epoch was used for weighting the individual data points ($1/\sigma^2$).
The fits give the separation rates $0.027 \pm 0.002$ mas~yr$^{-1}$ ($0.134 \pm 0.009\,c$, SW1--NE1), $0.022 \pm 0.002$ mas~yr$^{-1}$($0.108 \pm 0.012\,c$, SW2--NE1) and $-0.034 \pm 0.001$ mas~yr$^{-1}$($0.168 \pm 0.005\,c$, J1--NE1).
Positive velocities of SW1 and SW2 signify an advancing motion of both hotspots. The negative velocity of J1 indicates that the internal jet moves toward the terminal hotspot NE1. 
The separation of J1 at the first three epochs, 1995.266, 1995.395, and 1996.375, shows {\bf a strong} deviation from the general trend of the data points beyond 1997. During the fitting process, double uncertainties were assigned to these data points in order to reduce their weight and thus to avoid a bias induced by these large deviations. 
The separation of NE2--NE1 is about 1.2 mas, and the linear regression fit did not give a significant proper motion, because NE2 shows a significant transverse motion rather than a radial motion (Fig. \ref{fig:kin2D}). 

Liu et al. (2000) reported a first determination of the NE1--SW1 angular separation rate of $0.058\pm0.038$ mas~yr$^{-1}$, derived from 8.4-GHz data with only six epochs between 1994 and 1997.  {\bf Subsequently, } Luo et al. (2007) improved this value to $0.031\pm0.006$ mas/yr (5$\sigma$), on the basis of a longer time span (11 yr) and {\bf nine} epochs.
Compared to the previous 8.4-GHz work, the current proper motion measurements greatly improve the accuracy in the following ways: 
(1) The 15 GHz data provide a higher resolution (typically $0.8\times0.6$ mas$^{2}$), while the previous measurements at 8.4 GHz had a resolution ($1.3\times1.0$ mas$^{2}$) that is comparable with the separation between SW1 and SW2. 
At 8.4-GHz the positions of SW1 and SW2 are affected by the flux density variation of the two components (mainly SW1), which is also true for the northeast lobe where the positions of NE2 and J1 are affected by the flux density and structure variations of NE1.  At 15 GHz, SW1 is {\bf clearly} separated from SW2.
(2) The present work covers a time span of 13.6 yr, four times longer than in the earliest work by Liu et al. (2000), and $\sim$2.3 yr longer than that in Luo et al. (2007). 
(3) The data from 28 individual epochs give a better time sampling compared to the sparse sampling in previous works. 
(4) The 15-GHz VLBA data provide better and more uniform {\it u,v} coverage than the 8.4-GHz data.

According to the new measurement of $\mu=0.027\pm0.002$ mas~yr$^{-1}$ for SW1--NE1 14$\sigma$ detection), the kinematic age of OQ~208 is calculated as $255\pm17$ yr in the source rest frame. This age classifies OQ~208 as one of the youngest CSOs (\cite{pc03,gug05,gp09,an12a}).

Assuming equal advancing velocities for NE1 and SW1, the internal jet J1 is found to move 3.5 times faster (v$_{J1}$ = 0.23 c) than the terminal hotspots. A similar phenomenon has also been observed in other CSOs (e.g., B0710+439 and B2352+495: \cite{tay00}; J0132+5620: \cite{an12a}). 
While the radio lobes sweep the ambient ISM, a jet knot moves in an excavated channel with a relatively lower ISM density. 

The kinematic parameters of the OQ~208 jet derived above can be used to calculate the jet flow properties using the following the equations: 
\begin{equation}
\beta_{app} = \frac{\beta \sin\theta}{1 - \beta\cos\theta}
\end{equation} 
\begin{equation}
\Gamma = \frac{1}{\sqrt{1 - \beta^{2}} }
\end{equation} 
\begin{equation}
\delta = \frac{1}{\Gamma (1 - \beta\cos\theta)},
\end{equation} 
where, $\beta_{app}$ is the apparent speed of the jet component, $\beta$ is the intrinsic jet speed (both in units of $c$), $\theta$ is the inclination angle between the jet axis and the line of sight, $\Gamma$ is the bulk Lorentz factor, and $\delta$ is the Doppler boosting factor.
Taking the apparent speed of the jet knot J1 ($\beta_{app}=0.168\,c$) and the inclination angle ($\theta=80.8\degr$: Section 3.3) into account, we {\bf obtain} $\beta=0.166\,c$, $\Gamma=1.014$ and $\delta=1.013$. 
These calculations indicate a mildly relativistic flow in OQ~208. Relativistic beaming does not play a major role in determining the observed radiative and kinematic properties.  

\subsection{Sideways motion and frustrated jets}

\begin{figure*}
\centering
\includegraphics[angle=0,width=1.0\textwidth]{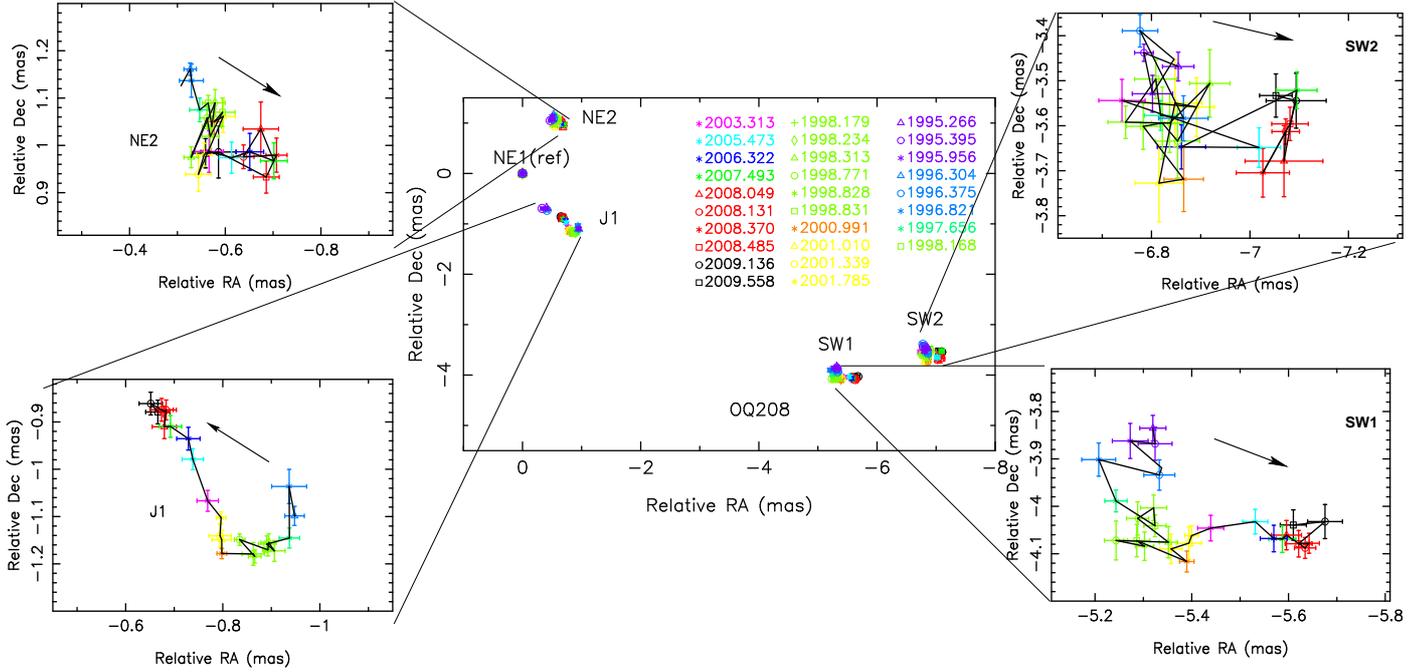}
\caption{Relative right ascension and declination of the VLBI components. The primary hotspot NE1 serves as a reference position. }
  \label{fig:kin2D}
\end{figure*}

In addition to the advancing motion of the hotspots, OQ~208 shows evidence of sideways motion. 
Figure~\ref{fig:kin2D} displays the two-dimensional plots of the VLBI components' relative locations. The primary hotspot NE1 is used as the reference at the (0,0) position. 
The southwest hotspot SW1 in general indicates an advancing motion along the NE1--SW1 line, although during the first half of the period 1996--2003 it shows a curved trajectory. Similar to SW1, the internal jet J1 shows a general motion toward NE1, but it makes a loop-like path during this same initial period.
This common feature of SW1 and J1 suggests that the reference NE1 may have followed a curved path between 1995 and 2003. 
If SW1 is used as the reference (plot is not shown here), NE1 indeed follows a bending trajectory, first to the north and then to the east, in a mirror-symmetric pattern with SW1--NE1. J1 follows a straight trajectory along the connecting line J1--SW1. 

SW2 shows a complex and disordered motion pattern (Fig. \ref{fig:kin2D}), but generally it moves to the southwest along the connecting line NE1--SW2 within the past fourteen years. \cite{wang03} measured a proper motion of SW2 (relative to NE1) as 0.032$\pm$0.020 mas~yr$^{-1}$ based on five-epoch 5-GHz data. Their measurement {\bf agrees with ours} ($\mu=0.022\pm0.002$ mas~yr$^{-1}$) with a much higher accuracy. 

The hotspot NE2 shows an apparent motion to the southwest, in an opposite direction of the jet advance. As discussed in Section 3.1, NE2 is likely a hotspot generated by the deflected jet hitting the wall of the surrounding ISM. While the hotspot NE1 is advancing faster than NE2, NE2 itself might be a stationary component. 

The wandering of the jet heads NE1 and SW1, as well as the disturbed lobe structure with deflected jet or double hotspots, provide the signature of an obstructed jet flow. 
According to the {\it frustration} model,  the advance motion of the radio lobes is confined by the surrounding dense ISM due to the intrinsically low jet power,  or the reduction or cessation of the jet power. 
A critical requirement for a CSO to evolve into medium-sized symmetric objects (MSO) is that the jet remains supersonic at the interface between the ISM and intergalactic medium (IGM). 
Through analytic modeling of the expansion of the hotspot and cocoon,  \cite{kk06} and \cite{kaw09} derived that the initial hotspot advance velocity should at least be 0.3$c$ so that the CSO can evolve beyond the ISM-IGM boundary of the host galaxy.
Jet flows with velocities below this threshold become subsonic before {\bf they reach} the ISM-IGM boundary and have a distorted morphology.

The present data show that the radio source of OQ~208 is still growing, although the advancing velocity is low. The radio power of OQ~208 is $P_{1.4GHz} = 10^{25.0}$ W~Hz$^{-1}$ without correction for the low-frequency absorption, placing it in the low-jet-power regime (\cite{ab12}).
A low-power CSO like OQ~208 can relatively easily develop hydrodynamic surface instabilities in the jet, {\bf which makes the jet more likely to loss momentum}. 
If a significant fraction of the jet momentum flux is dissipated, the jet cannot sustain {\bf the} supersonic laminar flow and forms a {\bf stagnating} standing shock ({\it i.e.}, the location of the hotspot) and becomes flaring and diffused beyond that point. Such a flared jet is found in the CSS quasar 3C~48, which has a compact and bright hotspot about 300 parsec from the central AGN (\cite{wil91,wor04,feng05,an10}).
The decrease of jet power can happen in any stage of the radio source evolution when the nuclear activity is reduced or terminated, or the jet experiences significant loss of kinetic energy because of jet-ISM interactions (\cite{ab12}).   
Frustrated sources may continue to grow but they do not have compact symmetric lobes. Eventually, frustrated CSOs and MSOs become radio relics at frequencies below a few hundred MHz. 

\section{Conclusions and summary} \label{sec4}

Multi-frequency multi-epoch radio images of OQ~208 with the highest angular resolution of $\sim 0.5$ mas present a typical CSO morphology with two compact mini-lobes along a position angle about $-126\degr$ and separated by $\sim$10 pc.
The flux density ratio of the structural components varies with the observing frequency.
At 8.4- and 15-GHz images, each lobe is resolved into two subcomponents, among which the brightest components are identified as the primary hotspots.
In the highest-resolution 15-GHz images, a knotty jet is detected between the NE and SW lobes. 
The core of OQ~208 cannot be securely identified from the present data.

The two primary hotspots NE1 and SW1 show significant flux density variations at 15 GHz of 62\% and 19\%, respectively.
The 15-GHz variability of the hotspots in OQ~208 may result from the balance between feeding from the central engine and adiabatic expansion loss.

{\bf The peak epoch of the NE1 flare in 1998.3 is earlier than that of the SW1 flare} by an estimated 5.00 yr, suggesting that NE1 is moving toward the observer and SW1 is receding. This light travel time difference between the two hotspots corresponds to a radial distance difference of about 1.53 pc.
Combining the projected and radial separation between NE1 and SW1, we estimate the inclination angle between the radio jet and the line of sight to be $\sim80.8\degr$. This value is {\bf higher} than the 70\degr opening angle generally assumed for the nuclear NLR, which suggest that the jet axis and the galaxy axis are not aligned.

Using the brightest hotspot NE1 as the reference, the relative proper motions {\bf were} estimated for SW1, SW2, and J1 {\bf to be} 0.027 mas~yr$^{-1}$, 0.022 mas~yr$^{-1}$, and $-$0.034 mas~yr$^{-1}$. 
The separation speed between SW1 and NE1 corresponds to a hotspot advancing speed of $0.065\,c$, assuming symmetric advancing and receding hotspot motions. The proper motion of the jet component J1 relative to the systemic center is about $-$0.047 mas~yr$^{-1}$ and corresponds to a velocity of 0.230 c, making the jet mildly relativistic.  
The internal jet knot moves significantly faster than the terminal hotspots because it encounters a lower density in the excavated jet channel than the hotspots within the lobes. 
The angular separation rate leads to an estimate of the kinematic age of about 255 yr, suggesting that OQ~208 is one of the youngest CSOs known.

The observed sideways motion of hotspots and disturbed lobe morphology are signatures of obstruction of the jet head by the surrounding ISM, although the overall radio source is still growing. 
During the early CSO stages of a radio source evolution, such disturbed lobe structures seem common (\cite{an12a}). A young radio source may experience intermittent jet power or several failed starts before entering into a continuous and steady growth phase. During each intermittent activity, the jets may impact different regions of the ISM, failing to make a breakthrough. The notion of intermittent energy feeding is supported by the detection of multiple hotspots and the fading lobe $\sim$30 mas ($\sim$40 pc) to the southwest in a different position angle from the main jet body. 
According to the evolution modeling of extragalactic radio sources (\cite{kb07}), a critical requirement for a CSO to evolve beyond the ISM--IGM boundary (typically 1--3 kpc) of the host galaxy is that the jet remains supersonic and maintains {\bf a} laminar flow. 
Compared to high-power high-velocity sources, low-power low-velocity CSOs such as OQ~208 can easily develop turbulent jet flow after losing a significant amount of {\bf their} momentum flux and kinetic energy during {\bf their} interactions with the ambient ISM.
Frustrated CSO radio sources with turbulent jet flow would not evolve into large-scale symmetric FR~II-type radio galaxies.

\section*{Acknowledgments}
\label{ack}
The authors thank the anonymous referee for helpful comments.
This work is supported in part by the National Basic Research Program of China (973 Program) under grant Nos. 2009CB24900 and 2013CB837901, the Science \& Technology Commission of Shanghai Municipality (06DZ22101), the China-Hungary Collaboration and Exchange Program funded by the International Cooperation Bureau of the Chinese Academy of Sciences (CAS) and the Strategic Priority Research Program (XDA04060700) of the CAS.
F.W. thanks the JIVE/ASTRON Summer Student Program and for the hospitality of the JIVE. S.F. was supported by the Hungarian Scientific Research Fund (OTKA K104539).
The authors thank Ger de Bruyn for discussions about low-frequency variability of OQ~208. 
This research has made use of the NASA/IPAC Extragalactic Database (NED), which is operated by the Jet Propulsion Laboratory, California Institute of Technology, under contract with the National Aeronautics and Space Administration.
The National Radio Astronomy Observatory is a facility of the National Science Foundation operated under cooperative agreement by Associated Universities, Inc.
The MOJAVE project is supported under National Science Foundation grant 0807860-AST and NASA-Fermi grant NNX08AV67G.
This research has made use of the United States Naval Observatory (USNO) Radio Reference Frame Image Database (RRFID).
The University of Michigan Radio Astronomy Observatory is supported by funds from the NSF, NASA, and the University of Michigan.

\clearpage
\small
\onecolumn
\begin{longtable}{ccccccccc}

\caption{Model-fitting parameters}\\
\hline\hline
Epoch    & $S$   & $\nu$ & Comp. &   $R$    & $\theta$   & $\theta_{maj}$ & $\theta_{min}$ & P.A.  \\
         &  (mJy)  & (GHz) &       & (mas)  &  (deg)     &      (mas)     &     (mas)      & (\degr) \\
(1)      &   (2)   & (3) &  (4)  &   (5)  &  (6)       &       (7)      &      (8)       & (9) \\
\hline

1995.266 & 659.65(46.2)&15.3 & NE1 & 0.0 (0.000)& $-$132.74 &       0.365     &      0.195     & 0.28 \\
         & 93.11 (6.5) &     & NE2 & 1.20(0.022)& $-$26.32  &       0.513     &      0.513    & $-$35.54  \\
         & 56.67 (4.0) &     & J1  & 0.80(0.042)& $-$148.46 &       1.023     &      1.023    & $-$112.62 \\
         & 39.74 (2.8) &     & SW1 & 6.56(0.056)& $-$125.95 &       0.883     &      0.883    & $-$149.04 \\
         & 16.68 (1.2) &     & SW2 & 7.68(0.068)& $-$116.73 &       0.527     &      0.527    & $-$107.10 \\
1995.395 & 712.48(49.9)&15.3 & NE1 & 0.0 (0.000)& 43.77     &       0.325     &      0.155     & $-$4.56 \\
         & 108.53(7.6)&      & NE2 & 1.15(0.035)& $-$24.11  &       0.549     &      0.549    & $-$155.56 \\
         & 42.50 (3.0)&      & J1  & 0.77(0.044)& $-$154.43 &       0.400     &      0.400    & $-$176.27\\
         & 39.84 (2.8)&      & SW1 & 6.58(0.070)& $-$126.00 &       0.855     &      0.855    & 141.34 \\
         & 16.36 (1.2)&      & SW2 & 7.61(0.026)& $-$116.87 &       0.410     &      0.410    & 10.74 \\
1995.956 & 723.35(50.6&15.3  & NE1 & 0.0 (0.000)& $-$147.26 &       0.484     &      0.243     & $-$2.25 \\
         & 75.48 (5.3)&      & NE2 & 1.24(0.037)& $-$24.26  &       0.594     &      0.594    & $-$155.56 \\
         & 21.71 (1.5)&      & J1  & 1.45(0.062)& $-$139.12 &       0.631     &      0.631    & $-$176.27 \\
         & 36.42 (2.5)&      & SW1 & 6.54(0.074)& $-$126.22 &       1.011     &      1.011    & $-$141.34 \\
         & 15.22 (1.1)&      & SW2 & 7.66(0.084)& $-$117.42 &       0.454     &      0.454    & $-$10.74 \\
1996.304 & 780.52(54.6)&15.3  & NE1 & 0.0 (0.000)& $-$132.59 &       0.431     &      0.199     & $-$9.72 \\
         & 72.79 (5.1) &      & NE2 & 1.28(0.022)& $-$24.40  &       0.589     &      0.589    & $-$155.56 \\
         & 17.82 (1.2) &      & J1  & 1.45(0.042)& $-$139.21 &       0.658     &      0.658    & $-$176.27 \\
         & 35.11 (2.5) &      & SW1 & 6.62(0.033)& $-$126.29 &       0.837     &      0.837    & 141.34 \\
         & 15.50 (1.1) &      & SW2 & 7.67(0.045)& $-$116.76 &       0.537     &      0.537    & 10.74 \\
1996.375 & 822.20(57.6)&15.3 & NE1 & 0.0 (0.000)& 33.49     &       0.371     &      0.179     & $-$11.31 \\
         & 86.51 (6.1)&      & NE2 & 1.19(0.041)& $-$25.63  &       0.573     &      0.573    & $-$59.04 \\
         & 53.55 (3.7)&      & J1  & 0.84(0.051)& $-$150.52 &       0.938     &      0.938    & $-$114.44\\
         & 37.58 (2.6)&      & SW1 & 6.63(0.062)& $-$126.42 &       0.830     &      0.830    & $-$176.19 \\
         & 14.55 (1.0)&      & SW2 & 7.58(0.074)& $-$116.57 &       0.430     &      0.430    & 180.00 \\
1996.821 & 767.11(53.7)&15.3 & NE1 & 0.0 (0.000)& $-$5.09   &       0.401     &      0.167     & $-$13.44 \\
         & 63.36 (4.4)&      & NE2 & 1.25(0.042)& $-$24.97  &       0.531     &      0.531    & $-$26.57 \\
         & 21.35 (1.4)&      & J1  & 1.40(0.076)& $-$137.89 &       0.810     &      0.810    & $-$162.90 \\
         & 28.77 (1.8)&      & SW1 & 6.51(0.073)& $-$126.84 &       0.648     &      0.648    & $-$120.96 \\
         & 21.92 (1.5)&      & SW2 & 7.74(0.110)& $-$117.57 &       0.771     &      0.771    & $-$122.01 \\
1997.656 & 813.14(56.9)&15.3 & NE1 & 0.0 (0.000)& 4.54      &       0.414     &      0.187     & $-$13.52 \\
         & 66.44 (4.6)&      & NE2 & 1.21(0.033)& $-$26.95  &       0.579     &      0.579    & $-$155.56 \\
         & 40.47 (2.8)&      & J1  & 1.48(0.031)& $-$140.71 &       0.355     &      0.355    & 176.19 \\
         & 33.54 (2.3)&      & SW1 & 6.59(0.040)& $-$127.26 &       0.706     &      0.706    & 141.34 \\
         & 15.20 (1.0)&      & SW2 & 7.70(0.068)& $-$117.68 &       0.686     &      0.686    & 10.74 \\
1998.168 & 864.91(60.5)&15.3 & NE1 & 0.0 (0.000)& $-$128.99 &       0.409     &      0.182     & $-$17.91 \\
         & 61.09 (4.3)&      & NE2 & 1.23(0.033)& $-$27.39  &       0.536     &      0.536    & $-$141.34 \\
         & 62.78 (4.4)&      & J1  & 1.46(0.020)& $-$142.44 &       0.212     &      0.212    & $-$118.61 \\
         & 30.89 (2.2)&      & SW1 & 6.68(0.045)& $-$127.20 &       0.558     &      0.558    & $-$156.80 \\
         & 16.04 (1.1)&      & SW2 & 7.65(0.089)& $-$117.18 &       0.591     &      0.591    & $-$145.31 \\
1998.179 & 654.88(45.8)&15.3 & NE1 & 0.0 (0.000)& 129.29    &       0.388     &      0.153     & $-$16.93 \\
         & 50.34 (3.5)&      & NE2 & 1.19(0.036)& $-$28.80  &       0.501     &      0.501    &  $-$71.57\\
         & 47.67 (3.3)&      & J1  & 1.48(0.030)& $-$142.30 &       0.190     &      0.190    & $-$156.04 \\
         & 25.22 (1.8)&      & SW1 & 6.66(0.047)& $-$126.95 &       0.652     &      0.652    & $-$140.71 \\
         & 13.81 (1.0)&      & SW2 & 7.64(0.070)& $-$118.03 &       0.799     &      0.799    & $-$118.61 \\
1998.234 & 731.95(51.2)&15.3 & NE1 & 0.0 (0.000)& $-$172.86 &       0.441     &      0.193     & $-$14.85 \\
         & 50.07 (3.5)&      & NE2 & 1.23(0.035)& $-$27.96  &       0.570     &      0.570    & $-$45.00 \\
         & 57.65 (4.0)&      & J1  & 1.46(0.025)& $-$142.24 &       0.250     &      0.250    & $-$151.93 \\
         & 25.95 (1.8)&      & SW1 & 6.65(0.042)& $-$127.28 &       0.594     &      0.594    & $-$142.13 \\
         & 16.05 (1.1)&      & SW2 & 7.72(0.074)& $-$117.73 &       0.791     &      0.791    & $-$124.99 \\
1998.313 & 909.06(63.6)&15.3 & NE1 & 0.0 (0.000)& 91.96     &       0.419     &      0.203     & $-$18.04 \\
         & 67.76 (4.7)&      & NE2 & 1.11(0.027)& $-$28.47  &       0.549     &      0.549    & $-$66.80 \\
         & 72.93 (5.1)&      & J1  & 1.47(0.017)& $-$142.66 &       0.240     &      0.240    & $-$104.93 \\
         & 40.75 (2.9)&      & SW1 & 6.73(0.038)& $-$127.29 &       0.775     &      0.775    & $-$154.98 \\
         & 13.36 (0.9)&      & SW2 & 7.76(0.085)& $-$116.87 &       0.605     &      0.605    & $-$135.00 \\
1998.771 & 853.14(59.7)&15.3 & NE1 & 0.0 (0.000)& 98.00     &       0.412     &      0.257     & $-$3.14 \\
         & 52.21 (3.7)&      & NE2 & 1.20(0.027)& $-$27.97  &       0.331     &      0.331    & $-$90.00 \\
         & 97.50 (6.8)&      & J1  & 1.42(0.016)& $-$144.00 &       0.254     &      0.254    & $-$61.39 \\
         & 30.00 (2.1)&      & SW1 & 6.64(0.048)& $-$127.83 &       0.619     &      0.619    & $-$142.43 \\
         & 19.25 (1.3)&      & SW2 & 7.76(0.094)& $-$117.94 &       0.856     &      0.856    & $-$105.52 \\
1998.828 & 872.47(61.1)&15.3 & NE1 & 0.0 (0.000)& 164.27    &       0.411     &      0.194     & $-$17.25 \\
         & 65.38 (4.6)&      & NE2 & 1.17(0.030)& $-$29.16  &       0.588     &      0.588    & $-$129.81 \\
         & 96.29 (6.7)&      & J1  & 1.47(0.019)& $-$143.83 &       0.210     &      0.210    & $-$119.75 \\
         & 34.01 (2.4)&      & SW1 & 6.69(0.035)& $-$127.61 &       0.559     &      0.559    & $-$153.44 \\
         & 16.98 (1.2)&      & SW2 & 7.71(0.091)& $-$117.78 &       0.711     &      0.711    & $-$144.46 \\
1998.831 & 861.22(60.3)&15.3 & NE1 & 0.0 (0.000)& 90.43     &       0.413     &      0.189     & $-$19.03 \\
         & 59.07 (4.1)&      & NE2 & 1.22(0.040)& $-$29.05  &       0.581     &      0.581    & $-$155.56 \\
         & 95.48 (6.7)&      & J1  & 1.46(0.033)& $-$143.81 &       0.188     &      0.188    & $-$176.27 \\
         & 34.67 (2.4)&      & SW1 & 6.67(0.043)& $-$127.61 &       0.614     &      0.614    & 141.34 \\
         & 16.32 (1.1)&      & SW2 & 7.68(0.061)& $-$117.97 &       0.611     &      0.611    & 10.74 \\
2000.991 & 756.35(52.9)&15.3 & NE1 & 0.0 (0.000)& $-$151.91 &       0.393     &      0.186     & $-$29.39 \\
         & 47.06 (3.3)&      & NE2 & 1.20(0.032)& $-$28.78  &       0.541     &      0.541    & $-$40.60 \\
         & 94.08 (6.6)&      & J1  & 1.42(0.016)& $-$145.89 &       0.178     &      0.178    & 173.66 \\
         & 39.42 (2.8)&      & SW1 & 6.78(0.027)& $-$127.37 &       0.481     &      0.481    & $-$165.96 \\
         & 14.67 (1.0)&      & SW2 & 7.81(0.083)& $-$118.44 &       0.723     &      0.723    & $-$112.62 \\
2001.010 & 753.72(52.8)&15.3 & NE1 & 0.0 (0.000)& 124.68    &       0.375     &      0.195     & $-$31.21 \\
         & 50.17 (3.5)&      & NE2 & 1.22(0.042)& $-$29.31  &       0.535     &      0.535    & $-$107.10 \\
         & 101.59(7.1)&      & J1  & 1.40(0.028)& $-$145.23 &       0.213     &      0.213    & $-$148.00 \\
         & 37.17 (2.6)&      & SW1 & 6.74(0.039)& $-$127.36 &       0.438     &      0.438    & $-$127.88 \\
         & 17.37 (1.2)&      & SW2 & 7.77(0.099)& $-$118.68 &       0.824     &      0.824    & $-$141.34 \\
2001.339 & 758.44(53.1)&15.3 & NE1 & 0.0 (0.000)& 98.70     &       0.344     &      0.179     & $-$31.03 \\
         & 57.32 (4.0)&      & NE2 & 1.09(0.045)& $-$30.08  &       0.538     &      0.538    & $-$155.56 \\
         & 106.86(7.5)&      & J1  & 1.39(0.033)& $-$145.15 &       0.244     &      0.244    & $-$176.27 \\
         & 41.45 (2.9)&      & SW1 & 6.76(0.043)& $-$127.09 &       0.475     &      0.475    & 141.34 \\
         & 16.03 (1.1)&      & SW2 & 7.71(0.111)& $-$117.40 &       0.684     &      0.684    & 10.74 \\
2001.785 & 750.57(52.5)&15.3 & NE1 & 0.0 (0.000)& $-$130.45 &       0.357     &      0.196     & $-$28.21 \\
         & 57.76 (4.0)&      & NE2 & 1.13(0.029)& $-$29.56  &       0.547     &      0.547    & $-$98.75 \\
         & 110.58(7.7)&      & J1  & 1.36(0.016)& $-$144.15 &       0.245     &      0.245    & $-$163.30 \\
         & 42.14 (2.9)&      & SW1 & 6.76(0.025)& $-$126.95 &       0.421     &      0.421    & $-$156.80 \\
         & 15.08 (1.1)&      & SW2 & 7.76(0.074)& $-$117.31 &       0.622     &      0.622    & $-$142.43 \\
2003.313 & 618.84(43.3)&15.3 & NE1 & 0.0 (0.000)& $-$57.71  &       0.395     &      0.215     & $-$34.13 \\
         & 54.58 (3.8)&      & NE2 & 1.14(0.053)& $-$29.84  &       0.616     &      0.616    & $-$174.91 \\
         & 125.72(8.8)&      & J1  & 1.32(0.031)& $-$144.22 &       0.328     &      0.328    & 180.00 \\
         & 42.00 (2.9)&      & SW1 & 6.78(0.046)& $-$126.64 &       0.524     &      0.524    & 153.92 \\
         & 16.24 (1.1)&      & SW2 & 7.61(0.095)& $-$117.74 &       0.774     &      0.774    & $-$167.73 \\
2005.473 & 438.03(30.7)&15.3 & NE1 & 0.0 (0.000)& $-$171.96 &       0.383     &      0.279     & $-$33.40 \\
         & 59.50 (4.2)&      & NE2 & 1.15(0.042)& $-$32.20  &       0.616     &      0.616    & $-$155.56 \\
         & 136.50(9.6)&      & J1  & 1.23(0.032)& $-$142.98 &       0.328     &      0.328    & 176.19 \\
         & 40.62 (2.8)&      & SW1 & 6.85(0.042)& $-$126.09 &       0.524     &      0.524    & 141.34 \\
         & 17.71 (1.2)&      & SW2 & 7.91(0.086)& $-$117.47 &       0.774     &      0.774    & 10.74 \\
2006.322 & 358.16(25.1)&15.3 & NE1 & 0.0 (0.000)& $-$103.38 &       0.369     &      0.232    & $-$26.88 \\
         & 50.20 (3.5)&      & NE2 & 1.18(0.049)& $-$33.44  &       0.638     &      0.638    & $-$111.80 \\
         & 135.62(9.5)&      & J1  & 1.18(0.035)& $-$142.06 &       0.341     &      0.341    & 168.69 \\
         & 32.98 (2.3)&      & SW1 & 6.90(0.046)& $-$126.14 &       0.473     &      0.473    & $-$146.31 \\
         & 20.30 (1.4)&      & SW2 & 7.77(0.096)& $-$117.99 &       0.992     &      0.992    & $-$106.70 \\
2007.493 & 297.76(20.8)&15.3 & NE1 & 0.0 (0.000)& 144.95    &       0.439     &      0.342     & $-$12.76 \\
         & 47.50 (3.3)&      & NE2 & 1.20(0.047)& $-$35.96  &       0.545     &      0.545    & $-$172.56 \\
         & 138.78(9.7)&      & J1  & 1.14(0.033)& $-$142.73 &       0.337     &      0.337    & $-$176.70 \\
         & 35.89 (2.5)&      & SW1 & 6.91(0.048)& $-$126.07 &       0.590     &      0.590    & $-$159.37 \\
         & 17.50 (1.2)&      & SW2 & 7.92(0.078)& $-$116.40 &       0.738     &      0.738    & $-$170.20 \\
2008.049 & 290.45(20.3)&15.3 & NE1 & 0.0 (0.000)& $-$172.66 &       0.498     &      0.405     & 55.75 \\
         & 42.41 (3.0)&      & NE2 & 1.24(0.068)& $-$33.10  &       0.680     &      0.680    & $-$126.87 \\
         & 117.24(8.2)&      & J1  & 1.14(0.036)& $-$143.25 &       0.380     &      0.380    & $-$168.69 \\
         & 34.20 (2.4)&      & SW1 & 6.91(0.054)& $-$125.97 &       0.536     &      0.536    & $-$158.20 \\
         & 15.50 (1.1)&      & SW2 & 7.97(0.164)& $-$117.49 &       0.643     &      0.643    & $-$126.87 \\
2008.131 & 328.61(23.0)&15.3 & NE1 & 0.0 (0.000)& 114.90    &       0.335     &      0.335    & $-$161.09 \\
         & 65.93 (4.6)&      & NE2 & 1.17(0.033)& $-$33.18  &       0.600     &      0.600    & 180.00 \\
         & 160.05(11.2)&     & J1  & 1.11(0.029)& $-$142.00 &       0.263     &      0.263    & 171.21 \\
         & 39.06 (2.7)&      & SW1 & 6.96(0.034)& $-$125.96 &       0.561     &      0.561    & 180.00 \\
         & 20.98 (1.5)&      & SW2 & 7.94(0.050)& $-$117.06 &       0.626     &      0.626    & 161.93 \\
2008.370 & 267.87(18.8)&15.3 & NE1 & 0.0 (0.000)& $-$161.26 &       0.429     &      0.355     & $-$8.18 \\
         & 42.41 (3.0)&      & NE2 & 1.21(0.042)& $-$35.83  &       0.456     &      0.456    & $-$172.56 \\
         & 122.81(8.6)&      & J1  & 1.11(0.016)& $-$142.25 &       0.277     &      0.277    & $-$176.70 \\
         & 31.06 (2.2)&      & SW1 & 6.96(0.04 )& $-$125.86 &       0.506     &      0.506    & $-$159.37 \\
         & 17.25 (1.2)&      & SW2 & 7.94(0.111)& $-$117.80 &       0.764     &      0.764    & $-$170.20 \\
2008.485 & 264.86(18.5)&15.3 & NE1 & 0.0 (0.000)& $-$123.69 &       0.382     &      0.382    & 174.91 \\
         & 49.90 (3.4)&      & NE2 & 1.16(0.043)& $-$36.31  &       0.553     &      0.553    & $-$172.56 \\
         & 135.60(9.4)&      & J1  & 1.10(0.034)& $-$142.38 &       0.320     &      0.320    & $-$176.70 \\
         & 35.01 (2.4)&      & SW1 & 6.95(0.045)& $-$125.97 &       0.615     &      0.615    & $-$159.37 \\
         & 17.00 (1.2)&      & SW2 & 7.94(0.069)& $-$116.92 &       0.699     &      0.699    & $-$170.20 \\
2009.136 & 216.27(15.3)&15.3 & NE1 & 0.0 (0.000)& 101.56    &       0.403     &      0.403    & 69.90 \\
         & 39.11 (3.0)&      & NE2 & 1.15(0.066)& $-$30.72  &       0.600     &      0.600    & $-$155.56 \\
         & 102.76(6.8)&      & J1  & 1.08(0.035)& $-$142.89 &       0.320     &      0.320    & 176.19 \\
         & 28.79 (2.0)&      & SW1 & 6.96(0.067)& $-$125.39 &       0.573     &      0.573    & 141.34 \\
         & 14.65 (1.1)&      & SW2 & 7.93(0.123)& $-$116.55 &       0.713     &      0.713    & 10.74 \\
2009.558 & 269.08(19.0)&15.3 & NE1 & 0.0 (0.000)& 118.00    &       0.463     &      0.36     & 48.97 \\
         & 62.23 (4.4)&      & NE2 & 1.13(0.041)& $-$29.61  &       0.752     &      0.752    & $-$47.49 \\
         & 109.22(7.7)&      & J1  & 1.10(0.036)& $-$142.84 &       0.326     &      0.326    & 33.74 \\
         & 34.94 (2.4)&      & SW1 & 6.91(0.042)& $-$125.76 &       0.673     &      0.673    & $-$86.25 \\
         & 18.48 (1.3)&      & SW2 & 7.89(0.058)& $-$116.61 &       0.704     &      0.704    & 54.08 \\
\hline
\label{tab:modfit}
\end{longtable}

\end{document}